\documentclass[12pt,english]{revtex4}
\usepackage[T1]{fontenc}
\usepackage[latin1]{inputenc}
\usepackage{babel}

\usepackage{textcomp}
\usepackage{amssymb}
\usepackage[unicode=true, pdfusetitle,
 bookmarks=true,bookmarksnumbered=false,bookmarksopen=false,
 breaklinks=false,pdfborder={0 0 1},backref=false,colorlinks=false]
 {hyperref}

\makeatletter
\@ifundefined{textcolor}{}
{%
 \definecolor{BLACK}{gray}{0}
 \definecolor{WHITE}{gray}{1}
 \definecolor{RED}{rgb}{1,0,0}
 \definecolor{GREEN}{rgb}{0,1,0}
 \definecolor{BLUE}{rgb}{0,0,1}
 \definecolor{CYAN}{cmyk}{1,0,0,0}
 \definecolor{MAGENTA}{cmyk}{0,1,0,0}
 \definecolor{YELLOW}{cmyk}{0,0,1,0}
 }

\makeatother

\begin{document}

\title{{\Large On the treatment of divergent integrals in perturbative quantum
field theory}}

\author{R. Trinchero}

\altaffiliation{Supported by CONICET}

\affiliation{Instituto Balseiro, 8400 Bariloche, Argentina}

\date{April 27, 2010}
\begin{abstract}
A simple integral that illustrates the concepts of regularization,
subtraction, renormalization and renormalization group employed in
perturbative quantum field theory(PQFT) is considered. 
\end{abstract}
\maketitle

\section{Introduction}

The description of processes where particles can be created and annihilated,
i.e. involving energies above the rest energy of the particles involved,
is well done in many cases in terms of relativistic quantum field
theories. The perturbative approximation to these quantum field theories
provides excellent agreement with experiment in the description of
fundamental interactions, such as for example the electroweak interactions\cite{chli}.
Probability amplitudes for different processes in the perturbative
approximation of quantum field theories are written in terms of Feynman
diagrams. Feynman rules associate to each diagram a certain contribution.
For diagrams which involve closed loops, the corresponding contribution
is in general given in terms of integrals that are not well defined,
they diverge. The renormalization program\cite{biblio} in perturbative
quantum field theory is the process that allows to obtain physical
results from these ill defined integrals. This program is both one
of the major and most historically controversial developments in the
subject. It is a fundamental tool in getting physical results and
at the same time involves a careful handling of ill defined integrals.
Nowadays renormalization is an important tool for any field-theorist.
From the point of view of learning renormalization the subject normally
presents some difficulties. These difficulties are mostly related
to the fact that in field theories describing nature the length and
subtlety of the calculations involved obscure the conceptual ideas
behind the renormalization procedure.

The purpose of this paper is to provide a pedagogical example of how
divergent integrals are dealt with in perturbative quantum field theory(PQFT).
The interest is focused not in how these integrals arise but on what
is done once you have them. As mentioned above in realistic examples
of this procedure the main ideas are obscured by the length of the
calculations involved. A example of these integrals is presented where
the concepts of regularization, subtraction, renormalization and renormalization
group are clearly illustrated even when the calculations involved
are very simple. Furthermore it permits to see in this particular
example different approaches to topics in this field, some of which
are considered in this paper. To the knowledge of the author, such
an example and its pedagogical treatment are not available in the
literature.

The paper is organized as follows. Section II presents the example,
analyses its superficial degree of divergence and introduces the concept
of regularization. Section III shows how finite parts independent
of the regularization employed can be obtained from the original integral.
Section IV presents the renormalization procedure and section V studies
its dependence on the subtraction point, leading to the concepts of
renormalization group transformation and renormalization group equations.

\section{The example}

\label{s1.3}

In PQFT the contribution of Feynman diagrams involving loops leads
to integrals of rational functions of the integration variables(internal
momenta) that depend on other variables(external momenta) and other
parameters(coupling constants, masses etc.). The momenta varying in
some $\mathbb{R}^{n}$ where $n$ is the dimension of space-time.
Next the following one-dimensional example is considered, \begin{equation}
I_{m}(q)=\int_{0}^{\infty}dp\ \frac{1}{p+q+m^{2}};\;\; q,m\in\mathbb{R}\ ,\ q>0\label{2.1}\end{equation}
here $p$($q$) play the role of internal(external) momenta and $m^{2}$
is a parameter. This integral is not well defined, it diverges. A
way to measure its degree of divergence is provided but what is usually
called superficial degree of divergence. This quantity measures the
behaviour of the integrand in its upper limit of integration. In order
to define it consider the following change of integration variable
$p=\lambda p'$ in the integral $I_{m}(q)$,\[
I_{m}(q)=\int_{0}^{\infty}dp'\;\lambda\frac{1}{\lambda p'+q+m^{2}}\]
and consider of the limit of the new integrand for $\lambda\to\infty$.
In this example it goes like $\lambda^{0}$ , in general the exponent
of $\lambda$ which is obtained in this way is what we refer to as
superficial degree of divergence(SDD) of the integral and we denote
by $\omega$. The integral of this limit is for this example,\[
I_{m}^{UV}=\int_{0}^{\infty}dp'\;\frac{1}{p'}\]
which is a logarithmically divergent integral. It is worth noting
that the handling of divergent integrals exemplified bellow is applicable
to integrands that for high integration momenta behave as an integer
power of this momenta. For those integrands the SDD is an integer.
In these cases if the integrand has no singular points in the interval
of integration and has SDD $\omega<0$ then the integral converges.

Going back to our example $I_{m}(q)$, it is noted that if it would
be well defined it would have the symmetry property, \begin{equation}
I_{m}(q)=I_{-m}(q)\ .\label{2.2}\end{equation}
Clearly, speaking about the symmetry of an object that is not well
defined is at least also not well defined. However symmetry is a fundamental
concept in physics, and in many cases it works as a guide in looking
for a way to make sense of the mathematical expressions that appear
in the description of physical theories. The treatment of divergent
integrals that appear in perturbative quantum field theory is not
an exception. As an example, the conservation of the electromagnetic
current in quantum electrodynamics(QED) is considered as a fundamental
physical requirement and the whole treatment of divergent integrals
in QED aims to maintain this symmetry.

\section{Regularization}

The process regularization of a integral consist in considering a
parameter dependent family of integrals which,
\begin{enumerate}
\item Is well defined for a continuous open interval of the parameter.
\item The original integral correspond to a limiting point of this interval(which
will be denoted as cut-off limit).
\end{enumerate}
For the example of the previous section the following regularization
is considered, \begin{equation}
I_{m}^{\Lambda}(q)=\int_{0}^{\Lambda}dp\frac{1}{p+q+m^{2}}\ ,\label{4.3}\end{equation}
 which leads to, \begin{eqnarray}
I_{m}^{\Lambda}(q) & = & \left[\ln(p+q+m^{2})\right]_{0}^{\Lambda}\nonumber \\
 & = & \ln\left[\frac{\Lambda+q+m^{2}}{q+m^{2}}\right]\ .\label{4.4}\end{eqnarray}
The original integral corresponds to the limit $\Lambda\to\infty$,
which in accordance with the value $\omega=0$, shows a logarithmic
divergence. This regularization preserves the symmetry (\ref{2.2}),
i.e.,\[
I_{m}^{\Lambda}(q)=I_{-m}^{\Lambda}(q)\]
such regularizations are called invariant.

Another regularization is provided by, \begin{eqnarray}
I_{m}^{M}(q) & = & \int_{0}^{\infty}dp\ \left\{ \frac{1}{p+q+m^{2}}-\frac{1}{p+q+mM}\right\} \nonumber \\
 & = & \left[\ln\frac{p+q+m^{2}}{p+q+mM}\right]_{0}^{\infty}\nonumber \\
 & = & -\ln\left[\frac{q+m^{2}}{q+mM}\right]=\ln{\left[\frac{q+mM}{q+m^{2}}\right]}\ .\label{4.5}\end{eqnarray}
The original integral corresponds to the limit $M\to\infty$, which
in accordance with the value $\omega=0$, shows a logarithmic divergence.
This regularization does not preserve the symmetry (\ref{2.2}),i.e.,\[
I_{m}^{M}(q)\neq I_{-m}^{M}(q)\]

\section{Subtraction}

Is it possible to separate a finite contribution from the regularized
integral in a way that does not depend on the regularization chosen?
It is clear that there are many ways to write the regularized integral
as the sum of two terms , one of which tends to a finite limit in
the cut-off limit and the other diverges in this limit. For example,
\begin{equation}
I_{m}^{\Lambda}(q)=I_{m,F}^{\Lambda}(q)+I_{m,D}^{\Lambda}(q)\ ,\label{4.6}\end{equation}
 with \begin{eqnarray}
I_{m,F}^{\Lambda}(q) & = & -\ln[q+m^{2}]\nonumber \\
I_{m,D}^{\Lambda}(q) & = & \ln[\Lambda+q+m^{2}]\ ,\label{4.7}\end{eqnarray}
 is a possibility, but it is also possible to take, \begin{eqnarray}
I_{m,F}^{\Lambda}(q) & = & -\ln\frac{q+m^{2}}{m^{2}+f(q)}\nonumber \\
I_{m,D}^{\Lambda}(q) & = & \ln\frac{\Lambda+q+m^{2}}{m^{2}+f(q)}\ ,\label{4.8}\end{eqnarray}
 for the moment there is no criteria to perform this separations.

In order to proceed derivatives of the integrand in (\ref{2.1}) will
be considered. Let ${\cal F}(p,q)$ denote this integrand,i.e.,\[
{\cal F}(p,q)=\frac{1}{p+q+m^{2}}\]
Then the integral of the derivative of the integrand respect to the
external momenta $q$ converges, indeed, \begin{eqnarray}
I{}_{m}^{(1)}(q) & \equiv & \int_{0}^{\infty}dp\ \frac{\partial{\cal F}(p,q)}{\partial q}\nonumber \\
 & = & \int_{0}^{\infty}dp\ \frac{-1}{(p+q+m^{2})^{2}}\nonumber \\
 & = & \left[\frac{1}{p+q+m^{2}}\right]_{0}^{\infty}=\frac{1}{q+m^{2}}\ .\label{4.10}\end{eqnarray}
where the first equality is just the definition of $I{}_{m}^{(1)}(q)$.
It is clear that since $I{}_{m}^{(1)}(q)$ is a convergent integral,
then any regularization of this integral would give in the cut-off
limit the same result obtained in (\ref{4.10}). Next consider the
Taylor expansion of the integrand $F(p,q)$ around $q=0$, i.e.,

\begin{equation}
{\cal F}(p,q)={\cal F}(p,0)+\left[\frac{d{\cal F}(p,q)}{dq}\right]_{q=0}q+\ldots+\left[\frac{d^{n}{\cal F}(p,q)}{dq^{n}}\right]_{q=0}q^{n}+\cdots\ .\label{4.11}\end{equation}
The only term in this expansion whose integral in $p$ does not converge
is the first one. The identification, \begin{equation}
I{}_{m}^{(1)}(q)=\frac{dI_{m}(q)}{dq}\label{4.12}\end{equation}
makes no sense because $I_{m}(q)$ is not well defined. However the
quantities, \begin{equation}
I_{m}^{(n)\Lambda}(q)=\frac{d^{n}I_{m}^{\Lambda}(q)}{dq^{n}}\label{4.13}\end{equation}
 \begin{equation}
I_{m}^{(n)M}(q)=\frac{d^{n}I_{m}^{M}(q)}{dq^{n}}\ .\label{4.14}\end{equation}
for any integer $n>0$ are well defined and satisfy,

\begin{equation}
\lim_{\Lambda\rightarrow\infty}I_{m}^{(n)\Lambda}(q)=\lim_{M\rightarrow\infty}I_{m}^{(n)M}(q)\ ,\label{4.15}\end{equation}
which simply reflects the independence on the regularization of convergent
integrals. Therefore the quantity,

\begin{eqnarray}
\tilde{I}_{m}(q) & = & \lim_{\Lambda\rightarrow\infty}[I_{m}^{\Lambda}(q)-I_{m}^{\Lambda}(0)]\nonumber \\
 & = & \lim_{\Lambda\rightarrow\infty}\left\{ \ln\left[\frac{\Lambda+q+m^{2}}{q+m^{2}}\right]-\ln\left[\frac{\Lambda+m^{2}}{m^{2}}\right]\right\} \nonumber \\
 & = & \ln\left[\frac{m^{2}}{q+m^{2}}\right]\label{4.16}\end{eqnarray}
 is finite and independent of the regularization employed, indeed,\begin{eqnarray}
\tilde{I}_{m}(q) & = & \lim_{M\rightarrow\infty}[I_{m}^{M}(q)-I_{m}^{M}(0)]\nonumber \\
 & = & \lim_{M\rightarrow\infty}\left\{ \ln\left[\frac{q+mM}{q+m^{2}}\right]-\ln\left[\frac{mM}{m}\right]\right\} \nonumber \\
 & = & \ln\left[\frac{m^{2}}{q+m^{2}}\right]\ .\label{4.17}\end{eqnarray}

\noindent This construction of finite parts is very similar to what
in the mathematical literature is called Hadamard´s finite parts.
See for example ref.\cite{balan} for an account of this subject. 

\noindent Next a series of remarks on the above construction are given,
\begin{enumerate}
\item What would happen is the SDD $\omega$ would equal some positive integer
$k$?. For this type of integrals taking the derivative of the integrand
respect to the external momenta reduces its SDD $\omega$ by 1. Therefore
the $k+1$ derivative of the integrand respect to the external momenta
would give a finite result when integrated. Thus subtracting to the
regularized integral the integral of the first $k$ terms of its Taylor
expansion as in (\ref{4.11}) leads to a finite quantity independent
of the regularization employed. More precisely if $K(q)$ denotes
the integral with SSD $\omega=k$ and $K^{\Lambda}(q)$ denotes the
corresponding regularized integral, then the quantity,\[
\widetilde{K}(q)=\lim_{\Lambda\rightarrow\infty}\left[K^{\Lambda}(q)-\left(K^{\Lambda}(0)+\frac{dK^{\Lambda}(q)}{dq}|_{q=0}\, q+\cdots+\frac{d^{k}K^{\Lambda}(q)}{dq^{k}}|_{q=0}\, q^{k}\right)\right]\]
is finite and independent of the regularization employed.
\item The quantity between parenthesis in the last equation is called the
subtraction. It is very important to note that the subtraction is
a polynomial in the external momenta $q.$ This implies that the Fourier
transform of the subtraction, i.e. its expression in coordinate space,
is a sum of derivatives of the delta function. Although not shown
in this work, the important point is that this same terms in coordinate
space would arise from adding to the Lagrangian of the field theory
a certain number of local terms in the fields and its derivatives.
This local terms are called counterterms and the coefficients in front
of them should be of higher order in the expansion parameter and dependent
on the regularization. Indeed temporal ordered products of fields,
from which these integrals arise, are just not well defined for coincident
space-time arguments, so the freedom of including counterterms can
be thought as arising from this indefinition.
\item The relation between the calculation of a given integral in two regularizations
is considered. If the result for the integral is known in one regularization
then it suffices to calculate the subtraction for the other regularization
in order to know the whole expression in this last regularization.
This is shown for the example of section \ref{s1.3}. Let $I_{m}^{Reg}(q)$
and $I_{m}^{Reg'}(q)$ denote the integral in two regularizations,
then,\[
I_{m}^{Reg}(q)=\tilde{I}_{m}(q)+I_{m}^{Reg}(0)\ ,\]
 and, \[
I_{m}^{Reg'}(q)=\tilde{I}_{m}(q)+I_{m}^{Reg'}(0)\ .\]
 thus, \[
I_{m}^{Reg'}(q)=I_{m}^{Reg}(q)-I_{m}^{Reg}(0)+I_{m}^{Reg'}(0)\ .\]

\item What can be said about the behaviour of the subtractions under the
transformation $m\to-m$? From eqs. (\ref{4.16}) it follows that
for the first regularization considered the subtraction is invariant,
however from (\ref{4.17}) it is clear that this is not the case for
the second regularization. The general result is that subtractions
are invariant if and only if the corresponding regularizations respect
the symmetry. At the level of the Lagrangian this implies that counterterms
are invariant under the symmetry if and only if the associated regularization
is invariant.
\item The above arguments suggest another way of getting a regularization
independent finite part of the integral. The original integral can
be replaced by one in which the first term in the Taylor's expansion
around $q=0$ of the integrand is subtracted to the integrand , i.e.
replacing , \[
I_{m}(q)=\int_{0}^{\infty}dp\ \frac{1}{p+q+m^{2}}\]
 by,\begin{equation}
\tilde{I}_{m}(q)=\int_{0}^{\infty}dp\ \left[\frac{1}{p+q+m^{2}}-\frac{1}{p+m^{2}}\right]\ .\label{4.18}\end{equation}
from which it is readily obtained, \begin{equation}
\tilde{I}_{m}(q)=\left[\ln\frac{p+q+m^{2}}{p+m^{2}}\right]_{0}^{\infty}=\ln\left[\frac{m^{2}}{q+m^{2}}\right]\ .\label{4.19}\end{equation}
which of course coincides with the result in (\ref{4.16}) and (\ref{4.17}).
This procedure allows to subtract without previously employing a regularization,
it was proposed in ref.\cite{zim}.
\end{enumerate}

\section{Renormalization}

As was shown in the previous section by means of the subtraction procedure
it is possible to separate in a regularization independent way a finite
part from the original divergent integral. However this does not solve
the problem of making physical sense of a theory were such divergent
integrals appear. In order to address this point it is necessary to
know how does the divergent integral enters in the physical theory.
Two cases will be considered here, the ones of masses and coupling
constants in quantum field theory(QFT), which correspond in general
to additive and multiplicative renormalization. In the first case
the integral appears in a physical quantity $E$ in the form, \begin{equation}
E=\mu_{0}+I_{m}(q)\ ,\label{4.20}\end{equation}
where $\mu_{0}$ is a parameter of the theory. In order to rewrite
(\ref{4.20}) in terms of the finite part $\tilde{I}_{m}(q)$ the
following identity is considered, \begin{equation}
E=\mu_{0}+\left[I_{m}^{Reg}(q)-I_{m}^{Reg}(0)\right]+I_{m}^{Reg}(0)\ ,\label{4.21}\end{equation}
defining the renormalized parameter $\mu_{R}$ by, \begin{equation}
\mu_{R}=\mu_{0}+I_{m}^{Reg}(0)\label{4.22}\end{equation}
and taking the cut-off limit leads to, \begin{equation}
E=\mu_{R}+\tilde{I}_{m}(q)\ .\label{4.23}\end{equation}
the parameter $\mu_{R}$ corresponds to a measurable physical quantity(the
mass for example) and therefore is given a finite value. Therefore
(\ref{4.22}) implies that the original parameter $\mu_{0}$ appearing
in the theory should depend on the regularization parameter such as
to cancel the contribution of $I_{m}^{Reg}(0)$. This is the key idea
in renormalization, bare quantities such as $\mu_{0}$ are not observable,
it does not matter what is their value. In other words, perturbation
theory is just an approximation, the unperturbed part of the Lagrangian
of the field theory under consideration has no physical reality, interactions
can not be switched of in nature they are always switched on.

For the case of coupling constant renormalization the integral appears
in a physical quantity $E'$ as follows,

\begin{equation}
E'=g_{0}\ (1+g_{0}\ I_{m}(q))\ .\label{4.24}\end{equation}
in terms of the finite part of $I_{m}(q)$ the last equation can be
written as, \begin{equation}
E'=g_{0}\,\left\{ 1+g_{0}\,\left[I_{m}^{Reg}(q)-I_{m}^{Reg}(0)\right]+g_{0}\, I_{m}^{Reg}(0)\right\} \ .\label{4.25}\end{equation}
defining the renormalized parameter $g_{R}$ by, \begin{equation}
g_{R}=g_{0}\,\left[1+g_{0}\, I_{m}^{Reg}(0)\right]\label{4.26}\end{equation}
 leads to, \begin{equation}
E'=g_{R}+g_{0}^{2}\,\tilde{I}_{m}(q)\ .\label{4.27}\end{equation}
Inverting (\ref{4.26}) and expanding for $g_{R}\ll1$, implies, \begin{equation}
E'=g_{R}\ \left[1+g_{R}\,\tilde{I}_{m}(q)\right]\ .\label{4.28}\end{equation}
This last expansion for $g_{R}\ll1$ is taken because the perturbative
expansion parameters are by definition the couplings constants. It
is important to note that the cut-off limit should be taken after
the renormalization.

It should be remarked that there is an alternative way to renormalize
the theory. This procedure is based on remark 2. of the last section.
The idea is to add local counterterms of higher order to the original
field theory in order to cancel the divergent contributions. The addition
of these counterterms being justified by the indefinition of temporal
ordered products for coincident spatial arguments\cite{bs}. This
scheme is more general than the previous one. This is so because it
allows to renormalize even when the regularization employed is not
invariant, you can not generate non-invariant counterterms by redefining
the original parameters appearing in the theory. This is so because
the original theory is invariant.

\section{Subtraction point and renormalization group}

Why was the expansion in (\ref{4.11}) done around $q=0$ and not
around other value $q=q_{s}$? In other words, is it possible to maintain
consistency considering the following subtraction point($q_{s})$
dependent finite part, \begin{equation}
\tilde{I}_{m}(q,q_{s})=\lim_{\Lambda\rightarrow\infty}\left[I_{m}^{Reg^{\Lambda}}(q)-I_{m}^{Reg^{\Lambda}}(q_{s})\right]\ ?\label{4.29}\end{equation}
 the one considered in (\ref{4.11}) is, \begin{equation}
\tilde{I}_{m}(q)=\tilde{I}_{m}(q,0)\ .\label{4.30}\end{equation}
The difference between both finite parts is a finite renormalization,
indeed, \begin{eqnarray}
\tilde{I}_{m}(q,q_{s}) & = & \lim_{\Lambda\rightarrow\infty}\left\{ I_{m}^{Reg^{\Lambda}}(q)-I_{m}^{Reg^{\Lambda}}(0)+\left[I_{m}^{Reg^{\Lambda}}(0)-I_{m}^{Reg^{\Lambda}}(q_{s})\right]\right\} \nonumber \\
 & = & \tilde{I}_{m}(q,0)+\Delta(0,q_{s})\ ,\label{4.31}\end{eqnarray}
where, \begin{eqnarray}
\Delta(0,q_{s}) & = & \lim_{\Lambda\rightarrow\infty}\left[I_{m}^{Reg^{\Lambda}}(0)-I_{m}^{Reg^{\Lambda}}(q_{s})\right]\nonumber \\
 & = & \lim_{\Lambda\rightarrow\infty}\left\{ \ln\left[\frac{\Lambda+m^{2}}{m^{2}}\right]-\ln\left[\frac{\Lambda+q_{s}+m^{2}}{q_{s}+m^{2}}\right]\right\} \nonumber \\
 & = & \ln\left[\frac{q_{s}+m^{2}}{m^{2}}\right]\label{4.32}\end{eqnarray}
is a finite quantity. In other words, a change in the subtraction
point just corresponds to the inclusion of finite renormalizations.
Such operations are known as renormalization group transformations.
The requirement that physical quantities do not depend on the choice
of subtraction point, i.e, \begin{equation}
\frac{dE}{dq_{s}}=0\ ,\label{4.33}\end{equation}
 leads to, \begin{equation}
\frac{d\mu_{R}}{dq_{s}}+\frac{d\tilde{I}_{m}(q,q_{s})}{dq_{s}}=0\ .\label{4.34}\end{equation}
which is usually called a renormalization group equation. Since the
second term does not vanish, as follows from eqs.(\ref{4.31}) and
(\ref{4.32}), then necessarily the parameter $\mu_{R}$ should depend
on the subtraction point $q_{s}$. Defining, \begin{equation}
\mu_{R}(q_{s})=\mu_{0}+I_{m}^{Reg}(q_{s})+const.=\mu_{R}(0)-\Delta(0,q_{s})+const.\label{4.35}\end{equation}
eq.(\ref{4.34}) holds, as follows from (\ref{4.31}). Proceeding
in a similar way for $E'$, leads to, \begin{equation}
g_{R}(q_{s})=g_{0}\ [1+g_{0}\, I_{m}^{Reg}(q_{s})]+const.\stackrel{g_{R}(0)\ll1}{=}g_{R}(0)[1-g_{R}(0)\ \Delta(0,q_{s})]+const.\ .\label{4.36}\end{equation}
the physical interpretation of the quantities $\mu_{R}(q_{s})$ and
$g_{R}(q_{s})$ is that they correspond to the value of masses and
coupling constants for processes involving momenta of the order of
$q_{s}.$ This dependence on the momentum scale is known as the \textquotedbl{}running\textquotedbl{}
of couping constants and masses.

\end{document}